\documentclass{article}
\usepackage{graphicx} 

\usepackage[top=30mm, bottom=30mm, left=30mm, right=30mm]{geometry}
\usepackage{graphicx}%
\usepackage{multirow}%
\usepackage{amsmath,amssymb,amsfonts}%
\usepackage{amsthm}%
\usepackage{mathtools}
\usepackage{mathrsfs}%
\usepackage[title]{appendix}%
\usepackage{xcolor}%
\usepackage{textcomp}%
\usepackage{manyfoot}%
\usepackage{booktabs}%
\usepackage{bm}%
\usepackage{natbib}%
\usepackage{here}%
\usepackage{enumitem}

\theoremstyle{plain}
\newtheorem{thm}{Theorem}
\newtheorem*{thm*}{Theorem}
\newtheorem{lem}{Lemma}

\title{A generalized ordinal quasi-symmetry model and its separability \\ 
for analyzing multi-way tables}
\author{Hisaya Okahara \thanks{Email : hisaya.okahara@gmail.com. Postal address : Yamazaki, Noda, 2788510, Chiba, Japan.} \and Kouji Tahata \and \\
Department of Information Sciences, Tokyo University of Science, Japan}
\date{\empty}

\begin{document}

\maketitle

\begin{abstract}
This paper addresses the challenge of modeling multi-way contingency tables for matched set data with ordinal categories. 
Although the complete symmetry and marginal homogeneity models are well established, they may not always provide a satisfactory fit to the data. 
To address this issue, we propose a generalized ordinal quasi-symmetry model that offers increased flexibility when the complete symmetry model fails to capture the underlying structure. 
We investigate the properties of this new model and provide an information-theoretic interpretation, elucidating its relationship to the ordinal quasi-symmetry model. 
Moreover, we revisit Agresti's findings and present a new necessary and sufficient condition for the complete symmetry model, proving that the proposed model and the marginal moment equality model are separable hypotheses. 
We demonstrate the practical application of our model through empirical studies on medical and public opinion datasets.
Comprehensive simulation studies evaluate the proposed model under various scenarios, including model's performance for multivariate normal data and asymptotic behavior.
It enables researchers to examine the symmetry structure in the data with greater precision, providing a more thorough understanding of the underlying patterns. 
This powerful framework equips researchers with the necessary tools to explore the complexities of ordinal variable relationships in matched data sets, paving the way for new discoveries and insights.
\end{abstract}

\section{Introduction} \label{intro}
Contingency tables play important roles in various fields, as they highlight relationships between categorical variables. 
Typically, the analysis of contingency tables is interested in whether the row and column variables are independent.
Pearson's chi-square and Fisher's exact tests are well known for testing the independence model.

Conversely, contingency tables with the same row and column classifications are often called square contingency tables.
Square contingency tables are obtained from, for example, matched pair data.
Typically, the analysis of square contingency tables considers the issue of symmetry rather than independence because observations tend to concentrate on or near the main diagonal.
Additionally, we obtained multi-way contingency tables from the matched data set with the same ordinal categories.
For example, Table \ref{tab:1}, taken directly from \citet{agresti2010Analysis}, shows the results of a three-period crossover study.
The purpose of the study was to compare the effectiveness of a placebo (treatment $A$), a low-dose analgesic (treatment $B$) and a high-dose analgesic (treatment $C$) in relieving severe uterine pain during menstruation, a condition known as dysmenorrhea.
The study randomly assigned participants to six possible sequences for administering the three treatments over three periods.
At the end of each period, the women rated the treatment as providing (1) no relief, (2) moderate relief, or (3) complete relief.
Table \ref{tab:1} summarizes the data combined across the six sequences.
\begin{table}[htbp]
	\centering
	\caption{Data from crossover study for treating dysmenorrhea \citep{agresti2010Analysis} \label{tab:1}}
	\begin{tabular}{ccccccccccc}
	\toprule
		&		&  \multicolumn{3}{c}{$B = 1$} & \multicolumn{3}{c}{$B = 2$} & \multicolumn{3}{c}{$B = 3$}	\\
	\cmidrule(lr){3-11}
	$A$	&	$C$	&	1	&	2	&	3	&	1	&	2	&	3	&	1	&	2	&	3	\\
	\midrule
	1	&		&	6	&	4	&	5	&	3	&	13	&	10	&	1	&	8	&	14	\\
	2	&		&	2	&	3	&	2	&	1	&	3	&	1	&	2	&	1	&	2	\\
	3	&		&	1	&	0	&	2	&	0	&	0	&	0	&	1	&	1	&	0	\\
	\bottomrule
	\end{tabular}
\end{table}


\noindent For such data, the structures of symmetry and marginal homogeneity are of particular interest, as they can provide insights into the underlying patterns and relationships within the data. 
Consequently, many statisticians have considered various symmetry and asymmetry models for analyzing square contingency tables, for example, \cite{bowker1948Test}, \cite{mccullagh1978Class}, \cite{agresti1983Simple} and \cite{tahata2020Separation}. 
These models range from simple and interpretable structures, such as complete symmetry, to more flexible and complex models that allow for different types of asymmetry. The choice of an appropriate model depends on the specific characteristics of the data and the research questions of interest.

Let $(X_{1},X_{2},\dots,X_{T})$ represent the $T$ responses in each matched set. 
A $T$-way contingency table with $r^{T}$ cells summarizes the possible outcomes when there are $r$ response categories. 
Each cell in the table can be represented by $\bm{i}=(i_{1},i_{2},\dots,i_{T})$, where the value of $X_{t}$ is $i_{t}$ for $t=1,2,\dots,T$.
For example, in Table \ref{tab:1}, we have a three-way contingency table where $T=3$ and $r=3$, i.e., with $3^3=27$ cells.

The probability of a cell $\bm{i}$ is given by $\pi_{\bm{i}} = \Pr(X_{1}=i_{1},X_{2}=i_{2},\dots,X_{T}=i_{T})$. We can also calculate the expected frequency of a cell $\bm{i}$ for a multinomial sample of size $n$ as $\mu_{\bm{i}}=n \pi_{\bm{i}}$.
Then, the probability of $X_{t}$ being equal to $j$ is given by
\[
	\Pr(X_{t}=j) = \pi_{+ \dots +j+ \dots +},
\]
where the subscript $j$ is in position $t$ and the symbol ``$+$'' means the corresponding sums.
$\Pr(X_{t}=j)$ represents the marginal distribution for $X_{t}$.

This $T$-way table satisfies \textit{marginal homogeneity} (MH) if
\[
	\Pr(X_{1}=j) = \Pr(X_{2}=j) = \dots = \Pr(X_{T}=j) \quad (j=1,\dots,r).
\]
It satisfies \textit{complete symmetry} (S) if
\[
	\pi_{\bm{i}} = \pi_{\bm{j}}
\]
for any permutation $\bm{j}=(j_{1},j_{2},\dots,j_{T})$ of $\bm{i}=(i_{1},i_{2},\dots,i_{T})$.
S implies MH, but the converse does not hold except when $T=r=2$.

For any $\bm{i}=(i_{1},i_{2},\dots,i_{T})$, $a$ is the minimum of $\bm{i}$, $b$ is the next smallest, $\dots$, and $m$ is the maximum.
The complete symmetry model is a log-linear model and one representation is
\[
	\log \pi_{\bm{i}} = \lambda_{ab \ldots m}.
\]
This notation reflects the permutation invariance of $\pi_{\bm{i}}$ in the subscript $\bm{i}$.

An $r^{T}$ table satisfies \textit{quasi-symmetry} (QS) if
\[
	\log \pi_{\bm{i}} = \lambda_{1}(i_{1}) + \lambda_{2}(i_{2}) + \dots + \lambda_{T}(i_{T}) + \lambda_{ab \ldots m}
\]
where $\lambda_{ab \ldots m}$ is defined as in the S model.
It has symmetric association and higher-order interaction terms, but allows each single-factor marginal distribution to have its own parameters.
For ordinal responses, a simpler model with quantitative main effects uses ordered scores $u_{1} \leq u_{2} \leq \dots \leq u_{r}$ where $u_{1}<u_{r}$.
The \textit{ordinal quasi-symmetry} (OQS) model is
\[
	\log \pi_{\bm{i}} = \beta_{1} u_{i_{1}} + \beta_{2} u_{i_{2}} + \dots + \beta_{T} u_{i_{T}} + \lambda_{ab \ldots m},
\]
where one can set $\beta_{T}=0$.
The complete symmetry is a special case of the OQS model where $\beta_{1}=\beta_{2}=\cdots=\beta_{T}$.
The OQS model is more parsimonious than the QS model.
Generally, (i) S implies OQS, but the converse does not hold, and (ii) S implies QS, but the converse does not hold.
When the OQS model does not fit the given data, we need to consider using a model with additional parameters. 
For instance, \citet{bhapkar1990Marginal} and \citet{tahata2011Generalized} have proposed such models. 
On the other hand, \citet{kateri1997Asymmetry}, \citet{kateri2007Class}, \citet{tahata2020Separation}, and \citet{tahata2022Advances} have suggested models based on $f$-divergence. 
Using the $f$-divergence, we can employ a model with the same number of parameters as the OQS model, but the structure of the model is different from the OQS model.

\citet{caussinus1965Contribution} pointed out that the QS model bridges the S and MH models in square contingency tables.
\citet{bishop2007Discrete} introduced the S model holds if and only if the QS and MH models hold for three-way contingency tables.
For $T$-way tables, \citet{bhapkar1990Marginal} prove the equivalence conditions of S for general order.
Additionally, the S model holds if and only if the OQS and MH models hold for $T$-way tables \citep{agresti2010Analysis}.
The OQS model implies the QS model, as mentioned earlier.
This implies that the MH model needs to be more relaxed in Agresti's findings.
In other words, we can replace the MH model with a model that has weaker constraints than the MH model.

In this paper, we have three objectives.
Firstly, we propose using a generalized OQS model when the S model fits the data poorly.
Secondly, we outline the properties of this model.
We also provide an information-theoretic interpretation of the OQS model from these properties.
Lastly, we revisit Agresti's findings and present a new necessary and sufficient condition for the S model.
This paper is organized as follows.
Section \ref{oqs} proposes the generalized OQS model.
Section \ref{iff} gives the equivalence condition of the S model.
Section \ref{gof} shows the properties of the goodness-of-fit test statistics.
Section \ref{exa} provides an example.
Section \ref{fin} summarizes this paper.

\section{Ordinal quasi-symmetry based on $f$-divergence} \label{oqs}
This section proposes the OQS model based on $f$-divergence, which we refer to as the OQS$[f]$ model throughout the paper. 
This model generalizes the concept of OQS model using $f$-divergence for $T$-way contingency tables. 
It extends the OQS$[f]$ model for two-way contingency tables proposed by \citet{kateri2007Class} to $T$-way tables.
Define
\[
	A(\bm{i}) = \{\bm{j}|\bm{j}~\mbox{is any permutation of}~(a,b,\dots,m)\}.
\]

We consider the case of $T=3$.
Consider an $r^{3}$ dimensional multinomial distribution with probability vector
\begin{multline*} 
	\bm{\pi}=(\pi_{111},\dots,\pi_{1r1},\pi_{211},\dots,\pi_{2r1},\dots,\\
		\pi_{r11},\dots,\pi_{rr1},\pi_{112},\dots,\pi_{rr2},\dots,\pi_{11r},\dots,\pi_{rrr})^{\top},
\end{multline*}
where ``$\top$'' denotes the transpose.
In addition, let $\bm{q}$ be a set of all cell probabilities equal to $1/r^{3}$, i.e., $\boldsymbol{q}=(1/r^{3},\dots,1/r^{3})^{\top}$. 
Consider all probability vectors $\boldsymbol{\pi}$ such that
\begin{align}
\label{c1}
	\underset{(s, t, u) \in A((i, j, k))}{\sum\sum\sum} \pi_{stu} = t_{abc}  \quad (i, j, k=1,\ldots,r).
\end{align}
Let $f$ be a strictly convex function on $(0, +\infty)$ with $f(1)=0$.
Then, the minimum value of the $f$-divergence proposed by \citet[Sec. 4]{csiszar2004Information}
\[
	I(\bm{\pi}:\bm{q})=\sum_{i=1}^{r}\sum_{j=1}^{r}\sum_{k=1}^{r} q_{ijk} f \left( \frac{\pi_{ijk}}{q_{ijk}} \right)
\]
is attained for $\pi_{ijk}^{S}=\alpha_{ijk}q_{ijk}$ where $\alpha_{ijk}=\alpha_{abc}$.
It should be noted that $\alpha_{ijk}$ is determined so that $\pi_{ijk}^{S}$ satisfies the restrictions and $\alpha_{ijk}=t_{abc}/(\sum_{(s, t, u) \in A((i, j, k))} q_{stu})$.
Then, we can obtain $\pi_{ijk}^{S}=t_{abc}/\#A((a,b,c))$, where the symbol ``\#'' denotes the number of elements in a set.
Here, we take $f(0)=\lim_{t \rightarrow 0}f(t)$, $0 \cdot f(0/0)=0$, and $0 \cdot f(a/0)=a \lim_{t \rightarrow \infty}[f(t)/t]$.

The $f$-divergence $I(\boldsymbol{\pi}:\boldsymbol{\pi}^{S})$ is minimized under the conditions that in addition to restriction (\ref{c1}),
\begin{align}
\label{c2}
	\sum_{i=1}^{r} u_{i} \pi_{i++} = \mu_{1}, \quad \sum_{j=1}^{r} u_{j} \pi_{+j+} = \mu_{2}, \quad \sum_{k=1}^{r} u_{k} \pi_{++k} = \mu_{3},
\end{align}
are given.
Then, note that (i) $I(\boldsymbol{\pi}:\boldsymbol{\pi}^{S})$ is strictly convex and (ii) the restrictions are linear equations.
The Lagrange function is written as
\begin{align*}
	L &=I(\boldsymbol{\pi}: \boldsymbol{\pi}^{S}) + \sum_{i=1}^{r}\sum_{j=1}^{r}\sum_{k=1}^{r} \phi_{ijk} \left( \underset{(s, t, u) \in A((i, j, k))}{\sum\sum\sum} \pi_{stu} - t_{abc} \right) \\
	&+ b_{1} \left( \sum_{i=1}^{r} u_{i} \pi_{i++} - \mu_{1} \right) + b_{2} \left( \sum_{j=1}^{r} u_{j} \pi_{+j+} - \mu_{2} \right) + b_{3} \left( \sum_{k=1}^{r} u_{k} \pi_{++k} - \mu_{3} \right). \\
\end{align*}
Equating the derivation of $L$ to 0 with respect to $\pi_{ijk}$ gives
\begin{align}
\label{c3}
	f^{\prime}\left( \frac{\pi_{ijk}}{\pi_{ijk}^{S}} \right) + \varphi_{ijk} + u_{i} b_{1} + u_{j} b_{2} + u_{k} b_{3} = 0,
\end{align}
where $f^{\prime}$ is denoted by $F$ and
\[
	\varphi_{ijk}= \underset{(s,t,u) \in A((i,j,k))}{\sum\sum\sum} \phi_{stu}.
\]
Let $\pi^{*}_{ijk}$ denote the solution satisfying (\ref{c1})--(\ref{c3}).
Furthermore, let $-b_{t}=\beta_{t}$ $(t=1,2,3)$ and $-\varphi_{ijk}=\psi_{ijk}$ $(i,j,k=1,\dots,r)$.
Then
\[
	\pi^{*}_{ijk} = \pi_{ijk}^{S} F^{-1} \left( \beta_{1}u_{i} + \beta_{2}u_{j} + \beta_{3}u_{k} + \psi_{abc} \right) \quad (i,j,k = 1,\dots,r),
\]
where $f^{\prime}$ is denoted by $F$.
The minimum value of $I(\boldsymbol{\pi} : \boldsymbol{\pi}^{S})$ is attained for $\pi^{*}_{ijk}$, where $\beta_{t}$ $(t=1,2,3)$ and $\psi_{abc}$ are determined so that $\pi^{*}_{ijk}$ satisfies restrictions (\ref{c1}) and (\ref{c2}).
Thus, we obtain the following lemma.
\begin{lem}
\label{lem1}
	Consider all probability vectors $\boldsymbol{\pi}$ such that in addition to restriction (\ref{c1}), the marginal moments (\ref{c2}) are given.
	Then the minimum value of $I(\boldsymbol{\pi} : \boldsymbol{\pi}^{S})$ is attained for 
	\[
		\pi^{*}_{ijk} = \pi_{ijk}^{S} F^{-1} \left( \beta_{1}u_{i} + \beta_{2}u_{j} + \beta_{3}u_{k} + \psi_{abc} \right) \quad (i,j,k = 1,\dots,r),
	\]
	where $\beta_{t}$ $(t=1,2,3)$ and $\psi_{abc}$ are determined so that $\pi^{*}_{ijk}$ satisfies the restrictions (\ref{c1}) and (\ref{c2}).
\end{lem}

Similarly to Lemma \ref{lem1}, we can generalize the result for $T \geq 4$.
For the $T$-way contingency tables, the restrictions (\ref{c1}) and (\ref{c2}) are replaced as follows:
\begin{align}
\label{c4}
	\underset{\bm{j} \in A(\bm{i})}{\sum \cdots \sum} \pi_{\bm{j}} = t_{ab \ldots m}  \quad (i_{t}=1,\ldots,r; t=1,\dots,T),
\end{align}
and
\begin{align}
\label{c5}
	\sum_{i_{1}=1}^{r} u_{i_{1}} \pi_{i_{1}+\cdots+} = \mu_{1},~  \sum_{i_{2}=1}^{r} u_{i_{2}} \pi_{+i_{2}+\cdots+} = \mu_{2},~  \dots, ~ \sum_{i_{T}=1}^{r} u_{i_{T}} \pi_{+ \cdots +i_{T}} = \mu_{T}.
\end{align}
Then, we can obtain the following theorem.
\begin{thm}
\label{thm1}
	Consider all probability vectors $\boldsymbol{\pi}$ such that in addition to the restriction (\ref{c4}), the marginal moments (\ref{c5}) are given.
	Then the minimum value of $I(\boldsymbol{\pi} : \boldsymbol{\pi}^{S})$ is attained for 
	\[
		\pi^{*}_{\bm{i}} = \pi_{\bm{i}}^{S} F^{-1} \left( \beta_{1} u_{i_{1}} + \beta_{2} u_{i_{2}} + \dots + \beta_{T} u_{i_{T}} + \psi_{ab \ldots m} \right),
	\]
	where $\beta_{t}$ $(t=1,2, \dots, T)$ and $\psi_{ab \dots m}$ are determined so that $\pi^{*}_{\bm{i}}$ satisfies the restrictions (\ref{c4}) and (\ref{c5}).
\end{thm}

Define
\[
	\pi_{\bm{i}}^{S} = \frac{1}{\#A(\bm{i})}\underset{\bm{j} \in A(\bm{i})}{\sum} \pi_{\bm{j}},
\]
where the symbol ``\#'' denotes the number of elements in a set.
It should be noted that $\pi_{\bm{i}}^{S}$ satisfies the symmetry structure.
We propose the OQS model based on the $f$-divergence as follows:
\[
	\pi_{\bm{i}} = \pi_{\bm{i}}^{S} F^{-1} \left( \beta_{1} u_{i_{1}} + \beta_{2} u_{i_{2}} + \dots + \beta_{T} u_{i_{T}} + \psi_{ab \ldots m} \right),
\]
where one can set $\beta_{T}=0$.
As introduced at the beginning of this section, we refer to this model as the OQS$[f]$ model. 
This notation emphasizes the model's dependence on the chosen $f$-divergence.
This model can be expressed as
\[
	F\left( \#A(\bm{i}) \pi_{\bm{i}}^{c} \right) = \beta_{1} u_{i_{1}} + \beta_{2} u_{i_{2}} + \dots + \beta_{T} u_{i_{T}} + \psi_{ab \ldots m},
\]
where
\[
	\pi_{\bm{i}}^{c} = \frac{\pi_{\bm{i}}}{\underset{\bm{j} \in A(\bm{i})}{\sum} \pi_{\bm{j}}}.
\]

It should be noted that the OQS[$f$] model with $f(x) = x \log (x)$ is equivalent to the OQS model.
Theorem \ref{thm1} explains the property of the OQS model.
It states that the OQS model can be perceived as a model closest to the S model in terms of the Kullback-Leibler divergence under the conditions (\ref{c4}) and (\ref{c5}).
This is a new interpretation of the OQS model from the point of view of information theory.

Consider the three-way contingency tables to easily discuss the properties of the OQS[$f$] model.
The OQS[$f$] model can express various structures by changing the function $f$.
$\pi_{ijk}^{c}$ is the conditional probability that the observation falls in the $(i,j,k)$ cell on condition that the observation falls in $(i,j,k), (j,i,k), \dots, (k,j,i)$ cells.
Under the OQS[$f$] model, 
\begin{align*}
F\left( \#A((i,j,k))\pi^{c}_{ijk} \right) - F\left( \#A((k,j,i))\pi^{c}_{kji} \right) &= (u_{i} - u_{k})\beta_{1},\\
F\left( \#A((i,j,k))\pi^{c}_{ijk} \right) - F\left( \#A((i,k,j))\pi^{c}_{ikj} \right) &= (u_{j} - u_{k})\beta_{2},\\
F\left( \#A((i,j,k))\pi^{c}_{ijk} \right) - F\left( \#A((j,i,k))\pi^{c}_{jik} \right) &= (u_{i} - u_{j})(\beta_{1}-\beta_{2}).
\end{align*}
Namely, the OQS[$f$] model indicates the difference structure between certain transformations based on the function $f$ for the conditional probabilities.
Specifically, it is important that a simple linear function of the parameters $\beta_{1}$, $\beta_{2}$, $\beta_{3}$ can express the difference structure.
Consider now three variables $U$, $V$, and $W$ having a joint normal distribution. 
\cite{yamamoto2007Decomposition} stated that if it is reasonable to assume underlying trivariate normal distribution with means $\mathrm{E}(U)=\mu_{1}$, $\mathrm{E}(V)=\mu_{2}$, $\mathrm{E}(W)=\mu_{3}$, variances $\mathrm{Var}(U)=\mathrm{Var}(V)=\mathrm{Var}(W)$, and correlations $\mathrm{Cor}(U, V)=\mathrm{Cor}(U, W)=\mathrm{Cor}(V, W)$, the OQS model with $\{u_{i}=i\}$ (which is called the LDPS model)  may be appropriate for the three-way contingency tables with ordinal categories. 
In other words, when this trivariate normal distribution can be assumed behind the contingency table, there is a high likelihood that the OQS[$f$] model will provide a good fit. 
This is because the OQS[$f$] model is more flexible than the LDPS model in terms of scores $u_{i}$ and function $f$.
Therefore, when the data exhibits a simple structure, as suggested by the OQS[$f$] model, there is a high likelihood that the model will provide a good fit. 
On the other hand, it should be noted that if the data does not exhibit such a simple structure, it may be necessary to apply a more relaxed QS model.

For example, when $f(x)=x \log (x)$, 
\begin{align*}
	\log \frac{\pi_{ijk}}{\pi_{kji}} &= (u_{i} - u_{k})\beta_{1},\\
	\log \frac{\pi_{ijk}}{\pi_{ikj}} &= (u_{j} - u_{k})\beta_{2},\\
	\log \frac{\pi_{ijk}}{\pi_{jik}} &= (u_{i} - u_{j})(\beta_{1}-\beta_{2}),
\end{align*}
Note that these equations for the case $f(x) = x \log (x)$ are identical to the general form presented above. 
This is because when $f(x) = x \log(x)$, $F(x) = \log(x) + 1$, and the denominators of the conditional probabilities cancel out in the logarithm.
On the other hand, when $f(x) = (1-x)^2$, the equations take a different form as shown below.
\begin{align*}
	\pi^{c}_{ijk} - \pi^{c}_{kji} &= (u_{i} - u_{k})\beta_{1}^{*},\\
	\pi^{c}_{ijk} - \pi^{c}_{ikj} &= (u_{j} - u_{k})\beta_{2}^{*},\\
	\pi^{c}_{ijk} - \pi^{c}_{jik} &= (u_{i} - u_{j})(\beta_{1}^{*}-\beta_{2}^{*}),
\end{align*}
where $\beta_{t}^{*}=\beta_{t}/(2 \#A((i,j,k)))$ $(t=1,2)$.
In other words, the OQS[$f$] model with $f(x)=x \log (x)$ (i.e., the OQS model) indicates the structure of the log ratio between the probabilities of two cells.
Additionally, the OQS[$f$] model with $f(x)=(1-x)^{2}$ indicates the structure of the difference between two conditional probabilities.
This model is referred to as the Pearsonian OQS (POQS) model.

\section{Necessary and sufficient condition for S model}
\label{iff}

This section presents a necessary and sufficient condition of the S model and revisits Agresti's findings.

Let $\{ u_{i} \}$ be a set of known scores $u_{1} \leq u_{2} \leq \dots \leq u_{r}$ where $u_{1}<u_{r}$ and $g(i) = u_{i}$ ($i=1,\dots,r$).
The marginal moment equality (ME) model is defined by
\[
	\mathrm{E}(g(X_{1}))=\mathrm{E}(g(X_{2}))= \dots = \mathrm{E}(g(X_{T})),
\]
where
\[
	\mathrm{E}(g(X_{t}))=\sum_{i_{t}=1}^{r} u_{i_{t}} \pi_{+\dots+i_{t}+\dots+}.
\]
It should be noted that the subscript $i_{t}$ is in position $t$.
MH implies ME, but the converse does not hold.
That is, the ME model is more relaxed than the MH model.

We consider the case of $T=3$.
If the S model holds, the OQS[$f$] and ME models also hold.
Assuming that both the OQS[$f$] and ME models hold, we can prove that the S model also holds.
To do so, we use the cell probabilities denoted by $\{ \tilde{\pi}_{ijk} \}$ that satisfy both the OQS[$f$] and ME structures simultaneously.
Since $\{ \tilde{\pi}_{ijk} \}$ satisfy the OQS[$f$] model,
\[
	F\left( \frac{\tilde{\pi}_{ijk}}{\tilde{\pi}_{ijk}^{S}} \right) = \beta_{1}u_{i} + \beta_{2}u_{j} + \beta_{3}u_{k} + \psi_{abc} \quad (i,j,k = 1,\dots,r),
\]
where
\[
	\tilde{\pi}_{ijk}^{S} = \frac{1}{\#A((i,j,k))}\underset{(s,t,u) \in A((i,j,k))}{\sum\sum\sum} \tilde{\pi}_{stu}.
\]
Then for any $(i,j,k)$,
\begin{align}
\label{p1}
	F\left( \frac{ \tilde{\pi}_{ijk}}{\tilde{\pi}_{ijk}^{S}} \right) - F\left( \frac{ \tilde{\pi}_{jik}}{\tilde{\pi}_{jik}^{S}} \right) = (u_{i} - u_{j})\beta_{1} + (u_{j} - u_{i})\beta_{2}.
\end{align}
The sum of the right side of equation (\ref{p1}) multiplied by $\tilde{\pi}_{ijk}$ gives
\[
	\sum_{i=1}^{r}\sum_{j=1}^{r}\sum_{k=1}^{r} \left( \beta_{1}(u_{i}-u_{j})\tilde{\pi}_{ijk} + \beta_{2}(u_{j}-u_{i})\tilde{\pi}_{ijk} \right)= 0,
\]
because $\{ \tilde{\pi}_{ijk} \}$ satisfies the structure of ME.
Therefore,
\begin{align}
\label{p2}
	\sum_{i=1}^{r} \sum_{j=1}^{r} \sum_{k=1}^{r} \tilde{\pi}_{ijk} \left( F\left( \frac{ \tilde{\pi}_{ijk}}{\tilde{\pi}_{ijk}^{S}} \right) - F\left( \frac{ \tilde{\pi}_{jik}}{\tilde{\pi}_{jik}^{S}} \right) \right) = 0.
\end{align}
Since the left side of equation (\ref{p2}) becomes
\[
	\sum_{k=1}^{r} \left( \sum_{i=1}^{r-1} \sum_{j=i+1}^{r} \left( \tilde{\pi}_{ijk} - \tilde{\pi}_{jik} \right) \left( F\left( \frac{ \tilde{\pi}_{ijk}}{\tilde{\pi}_{ijk}^{S}} \right) - F\left( \frac{ \tilde{\pi}_{jik}}{\tilde{\pi}_{jik}^{S}} \right) \right) \right)
\]
and the function $F$ is a monotonically increasing function, we can obtain $\tilde{\pi}_{ijk} = \tilde{\pi}_{jik}$ for $i<j$ and $k=1,\dots,r$.
In a similar way, we can get $\tilde{\pi}_{ijk} = \tilde{\pi}_{abc}$.
That is, $\{ \tilde{\pi}_{ijk} \}$ satisfies the structure of the S model.

This discussion can be generalized in the case of $T \geq 4$.
We can obtain the following theorem.
\begin{thm}
	\label{thm2}
	The S model holds if and only if both the OQS[$f$] and ME models hold.
\end{thm}

According to \cite{bhapkar1990Marginal}, the S model holds if and only if both the QS and the MH models hold.
On the other hand, \cite{agresti2010Analysis} states that the S model holds if and only if both the OQS and MH models hold.
These two findings clearly show that the QS model can be replaced by the OQS model because the OQS model implies the QS model.
This means that requiring both OQS and MH to hold might be redundant.
Therefore, Agresit's research suggests that the MH model needs to be relaxed.
Theorem \ref{thm2} demonstrates that the ME model can substitute the MH model, and the condition in which both OQS and ME hold does not have any redundancy.

We note that Theorem \ref{thm2} includes the results of \cite{saigusa2015Orthogonal}.
For two-way contingency tables, \cite{kateri2007Class} described that the S model holds if and only if both the OQS[$f$] model and the MH model \citep{stuart1955Test} hold, while \cite{saigusa2015Orthogonal} showed that the S model can be separated into the OQS[$f$] model and the ME model.

\section{Goodness-of-fit test}
\label{gof}

Let $n_{\bm{i}}$ denote the observed frequency in the $\bm{i}=(i_{1},\dots ,i_{T})$th cell of the $r^T$ $(T\geq 2)$ table $(i_{k}=1,\dots ,r;k=1,\dots ,T)$, where $n=\sum \dots \sum n_{i_{1}\dots i_{T}}$, and let $m_{\bm{i}}$ denote the corresponding expected frequency.
Assume that $\{n_{\bm{i}}\}$ has a multinomial distribution.
The maximum likelihood estimates (MLEs) of expected frequencies $\{m_{\bm{i}}\}$ under the S, OQS and QS models could be obtained by using the generalized iterative scaling for log-linear models \citep{darroch1972Generalized}. Alternatively, \texttt{R} program \texttt{glm} can also be convenient for these models.
On the other hand, those methods can not be applied to the OQS$[f]$ and ME models, and therefore, the Newton-Raphson method is applied to the log-likelihood equations.
\cite{lang2004Multinomialpoisson} presented the unified theory of maximum likelihood inference for a broad class of contingency table models and further implemented the results in \texttt{R} (that is, \texttt{mph.fit}).
Thus, \texttt{R} program \texttt{mph.fit} is useful to fit these models \citep{lang2004Multinomialpoisson, lang2005Homogeneous}.

Let $G^{2}(\mathrm{M})$ denote the likelihood ratio statistic for testing the goodness-of-fit of model M:
\[
	G^{2}(\mathrm{M})=2 \sum_{i_{1}=1}^{r} \dots \sum_{i_{T}=1}^{r} n_{i_{1}\dots i_{T}} \log \left( \frac{n_{i_{1}\dots i_{T}}}{\hat{m}_{i_{1}\dots i_{T}}} \right),
\]
where $\hat{m}_{i_{1}\dots i_{T}}$ is the maximum likelihood estimate of the expected frequency $m_{i_{1}\dots i_{T}}$ under the model M.
The goodness of fit of each model can be tested using the likelihood ratio chi-squared statistic with the corresponding degrees of freedom (df).
The number of df for the S model is 
\[
	r^{T} - \left( \begin{array}{cc}
	r + T - 1 \\
	T 
 \end{array} \right).
\]
Additionally, the numbers of df for the OQS[$f$] and ME models are
\[
	r^{T} - \left( \begin{array}{cc}
	r + T - 1 \\
	T 
 \end{array} \right) - (T-1) \quad \mbox{and} \quad T-1,
\]
respectively.

Generally, for any set of models, suppose that model M$_{3}$ holds if and only if both models M$_{1}$ and M$_{2}$ hold.
For example, \citet{aitchison1962Largesample}, \citet{darroch1963Testing}, and \citet{tomizawa2007Analysis} discussed the properties of test statistics for model separation.
As described in \cite{aitchison1962Largesample}, (i) when the following asymptotic equivalence holds:
\begin{equation}
\label{orth}
	G^{2}(\mbox{M}_{3}) \simeq G^{2}(\mbox{M}_{1}) + G^{2}(\mbox{M}_{2}),
\end{equation}
where the number of df for M$_{3}$ equals the sum of numbers of df for M$_{1}$ and M$_{2}$, if both M$_{1}$ and M$_{2}$ are accepted (at the $\alpha$ significance level) with high probability, then M$_{3}$ would be accepted; however (ii) when (\ref{orth}) does not hold, such an incompatible situation that both M$_{1}$ and M$_{2}$ are accepted with high probability but M$_{3}$ is rejected is quite possible.
It should be noted that the number of df for S equals the sum of the numbers of df for OQS[$f$] and ME.
It is essential for simultaneous modeling of joint and marginal distributions, as mentioned in \citet{lang1994Simultaneously}.
Thus, this section shows the relationship between the goodness-of-fit statistics related to Theorem \ref{thm2}.

Consider the case of $T=3$.
The OQS[$f$] model may be expressed as 
\begin{align}
\label{ort1}
	F\left( \frac{ \pi_{ijk}}{\pi_{ijk}^{S}} \right) = (u_{i}-u_{k}) \beta_{1} + (u_{j}-u_{k}) \beta_{2} + \lambda_{abc}.
\end{align}
Let
\begin{multline*}
	F \left( \frac{\bm{\pi}}{\bm{\pi}^{S}} \right)=(F_{111},\dots, F_{1r1}, F_{211},\dots, F_{2r1},\dots, F_{r11},\dots,F_{rr1},F_{112},\dots,F_{rr2},\dots,F_{11r},\dots,F_{rrr})^{\top},
\end{multline*}
where
\[
	F_{ijk} = F\left( \frac{ \pi_{ijk}}{\pi_{ijk}^{S}} \right).
\]
Then the OQS[$f$] model can be expressed as
\[
	F \left( \frac{\bm{\pi}}{\bm{\pi}^{S}} \right) = \bm{X \beta} = (\bm{x}_{1},\bm{x}_{2},\bm{X}_{123}) \bm{\beta},
\]
where $\bm{X}$ is the $r^{3} \times K$ matrix with $K=2 + (r+2)(r+1)r/6$. The components of $\bm{X}$ are as follows:
\begin{align*}
	\bm{x}_{1} &= \bm{1}_{r} \otimes \bm{J}_{r} \otimes \bm{1}_{r} - \bm{J}_{r} \otimes \bm{1}_{r^{2}}: \ r^{3} \times 1 \ {\rm vector},\\
	\bm{x}_{2} &= \bm{1}_{r^{2}} \otimes \bm{J}_{r} - \bm{J}_{r} \otimes \bm{1}_{r^{2}}: \ r^{3} \times 1 \ {\rm vector},
\end{align*}
and the $\bm{X}_{123}$ is $r^{3} \times (r+2)(r+1)r/6$ matrix with elements $0$ or $1$ determined from \eqref{ort1}. Here, $\bm{1}_{s}$ is an $s \times 1$ vector of ones, $\bm{J}_{r}=(u_{1},\dots,u_{r})^{\top}$, and $\otimes$ denotes the Kronecker product.
Additionally,
\[
	\bm{\beta} = (\beta_{1}, \beta_{2}, \bm{\lambda}_{123})^{\top}, 
\]
where
\[
	\bm{\lambda}_{123} = (\lambda_{111},\dots,\lambda_{11r},\lambda_{122},\dots,\lambda_{12r},\dots,\lambda_{222},\dots,\lambda_{22r},\dots,\lambda_{(r-1)rr},\lambda_{rrr}).
\]
It should be noted that (i) $\bm{\lambda}_{123}$ is the $1 \times (r+2)(r+1)r/6$ vector, (ii) $\bm{X}_{123} \bm{1}_{(r+2)(r+1)r/6} = \bm{1}_{r^{3}}$ and (iii) the matrix $\bm{X}$ has full column rank which is $K$.

Similarly to \cite{lang1994Simultaneously}, we denote the linear space spanned by the columns of the matrix $\bm{X}$ as $S(\bm{X})$ with dimension $K$.
$S(\bm{X})$ is a subspace of $\mathbb{R}^{r^{3}}$.
Let $\bm{U}$ be an $r^{3} \times d_{1}$ full column rank matrix, where $d_{1} = r^{3} - K =-2+r(r-1)(5r+2)/6$, such that the linear space spanned by the columns of $\bm{U}$, denoted by $S(\bm{U})$, is the orthogonal complement of the space $S(\bm{X})$.
Thus, $\bm{U}^{\top}\bm{X}=\bm{O}_{d_{1},K}$ where $\bm{O}_{d_{1},K}$ is the $d_{1} \times K$ zero matrix.
Therefore the OQS[$f$] model is expressed as
\[
	 \bm{h}_{1}(\bm{\pi}) = \bm{U}^{\top} F \left( \frac{\bm{\pi}}{\bm{\pi}^{S}} \right) = \bm{0}_{d_{1}}
\]
where $\bm{0}_{d_{1}}$ is the $d_{1} \times 1$ zero vector.

The ME model may be expressed as
\[
	\bm{h}_{2}(\bm{\pi}) = \bm{W}\bm{\pi} = \bm{0}_{d_{2}}
\]
where $d_{2} = 2$ and $\bm{W} = \left( \bm{x}_{1}, \bm{x}_{2} \right)^{\top}$.
It should be noted that $\bm{W}$ is the $d_{2} \times r^{3}$ matrix.

Thus, the column vectors of $\bm{W}^{\top}$ belong to the space $S(\bm{X})$, i.e., $S(\bm{W}^{\top}) \subset S(\bm{X})$.
Therefore, $\bm{WU}=\bm{O}_{d_{2},d_{1}}$.
From Theorem \ref{thm2}, the S model can be expressed as $\bm{h}_{3}(\bm{\pi}) = \bm{0}_{d_{3}}$ where $\bm{h}_{3}(\bm{\pi}) = (\bm{h}_{1}^{\top}(\bm{\pi}), \bm{h}_{2}^{\top}(\bm{\pi}))^{\top}$ and $d_{3}=d_{1}+d_{2}=r(r-1)(5r+2)/6$.

Let $\bm{H}_{s}(\bm{\pi})$, $s=1,2,3$, denote the $d_{s} \times r^{3}$ matrix of partial derivatives of $\bm{h}_{s}(\bm{\pi})$ with respect to $\bm{\pi}$, i.e., 
\[
	\bm{H}_{s}(\bm{\pi})= \frac{\partial \bm{h}_{s}(\bm{\pi})}{\partial \bm{\pi}^{\top}},
\]
and define $\bm{\Sigma}(\bm{\pi}) = \bm{D}(\bm{\pi}) - \bm{\pi \pi}^{\top}$ where $\bm{D}(\bm{\pi})$ denotes a diagonal matrix with $i$th component of $\bm{\pi}$ as $i$th diagonal component.
Let $\bm{p}$ denote $\bm{\pi}$ with $\pi_{ijk}$ replaced by $p_{ijk}$, where $p_{ijk}=n_{ijk}/n$.
Then $\sqrt{n}(\bm{p}-\bm{\pi})$ has asymptotically a normal distribution with mean $\bm{0}_{r^{3}}$ and covariance matrix $\bm{\Sigma}(\bm{\pi})$.
Using the delta method,  $\sqrt{n}(\bm{h}_{3}(\bm{p})-\bm{h}_{3}(\bm{\pi}))$ has asymptotically a normal distribution with mean $\bm{0}_{d_{3}}$ and covariance matrix   
\[
	\bm{H}_{3}(\bm{\pi}) \bm{\Sigma} (\bm{\pi}) \bm{H}_{3}^{\top} (\bm{\pi}) = \left[
	\begin{array}{cc}
		\bm{H}_{1}(\bm{\pi}) \bm{\Sigma}(\bm{\pi}) \bm{H}_{1}^{\top} (\bm{\pi}) & \bm{H}_{1}(\bm{\pi}) \bm{\Sigma}(\bm{\pi}) \bm{H}_{2}^{\top}(\bm{\pi}) \\
		\bm{H}_{2}(\bm{\pi}) \bm{\Sigma}(\bm{\pi}) \bm{H}_{1}^{\top}(\bm{\pi}) & \bm{H}_{2}(\bm{\pi}) \bm{\Sigma}(\bm{\pi}) \bm{H}_{2}^{\top}(\bm{\pi}) \\
	\end{array}
	\right].
\]  
Then
\[
	\bm{H}_{1}(\bm{\pi}) = \bm{U}^{\top} \frac{\partial}{\partial \bm{\pi}^{\top}} F \left( \frac{\bm{\pi}}{\bm{\pi}^{S}} \right),
\]
where the element of $\dfrac{\partial}{\partial \bm{\pi}^{\top}} F \left( \dfrac{\bm{\pi}}{\bm{\pi}^{S}} \right)$ is given as
\[
	\frac{\partial}{\partial \pi_{stu}} F \left( \frac{\pi_{ijk}}{\pi_{ijk}^{S}} \right) = \frac{1}{\pi_{ijk}^{S}} \left( 1 - \pi_{ijk}^{c} \right) F^{\prime} \left( \frac{\pi_{ijk}}{\pi_{ijk}^{S}} \right),
\]
for $(s,t,u) = (i,j,k)$,
\[
	\frac{\partial}{\partial \pi_{stu}} F \left( \frac{\pi_{ijk}}{\pi_{ijk}^{S}} \right) = - \frac{1}{\pi_{ijk}^{S}} \pi_{ijk}^{c} F^{\prime} \left( \frac{\pi_{ijk}}{\pi_{ijk}^{S}} \right),
\]
for $(s,t,u) \neq (i,j,k)$ and $(s,t,u) \in A((i,j,k))$, and
\[
	\frac{\partial}{\partial \pi_{stu}} F \left( \frac{\pi_{ijk}}{\pi_{ijk}^{S}} \right) = 0,
\]
for the others. 

Additionally,
\[
	\bm{H}_{2}(\bm{\pi})=\bm{W}.
\]
 
Under the hypothesis $\bm{h}_{3}(\bm{\pi})=\bm{0}_{d_{3}}$, since $\bm{H}_{1}(\bm{\pi}) \bm{\Sigma}(\bm{\pi}) \bm{H}_{2}^{\top} (\bm{\pi}) = \bm{O}_{d_{1},d_{2}}$, we obtain that
\[
	\Delta_{3} = \Delta_{1} + \Delta_{2},
\]
where
\[
	\Delta_{s} = n \bm{h}_{s}(\bm{p})^{\top} \left( \bm{H}_{s}(\bm{p}) \bm{\Sigma}(\bm{p}) \bm{H}_{s}^{\top}(\bm{p}) \right)^{-1} \bm{h}_{s}(\bm{p}).
\]

The test statistic $\Delta_{s}$ asymptotically follows a chi-squared distribution with $d_{s}$ df, where $s=1,2,3$.
Specifically, (i) $\Delta_{1}$ is the test statistic for the OQS[$f$] model, (ii) $\Delta_{2}$ is that for the ME model and (iii) $\Delta_{3}$ is that for the S model.
From the asymptotic equivalence of the test statistic and the likelihood ratio statistic \cite[Sec. 6e. 3]{rao1973Linear}, under the S model, we obtain
\[
	G^{2}(\mbox{S}) \simeq G^{2}(\mbox{OQS[$f$]}) + G^{2}(\mbox{ME}).
\]

This discussion can be generalized to the case of $T \geq 4$, leading to the following theorem.

\begin{thm}
\label{thm3}
	 Under the S model, $G^{2}(\mbox{S})$ is asymptotically equivalent to the sum of $G^{2}(\mbox{OQS}[f])$ and $G^{2}(\mbox{ME})$.
\end{thm}

The incompatible situation described above would not arise because the OQS[$f$] and ME models are separable hypotheses from Theorems \ref{thm2} and \ref{thm3}.
It should be noted that Theorem \ref{thm3} was given by \cite{saigusa2015Orthogonal} for two-way contingency tables.
We point out that the condition ``under the S model'' is unnecessary when we use $f(x) = x \log (x)$.
That is, the test statistic $G^{2}(\mbox{S})$ is asymptotically equivalent to the sum of $G^{2}(\mbox{OQS})$ and $G^{2}(\mbox{ME})$.
This is related to \cite{tahata2011Generalized}.

The hypothesis $H_{0}:\beta_{1}=\beta_{2}=\beta_{3}$ can be tested using
\[
	G^{2}({\rm S}|{\rm OQS}[f])=G^{2}({\rm S})-G^{2}({\rm OQS}[f]),
\]
because the S and OQS[$f$] models are hierarchical.
The numbers of df for the chi-squared distribution differ between the two models.
From Theorem \ref{thm2}, we can see that
\[
	G^{2}({\rm S}|{\rm OQS}[f])=G^{2}({\rm ME}|{\rm OQS}[f]).
\]
That is, the conditional test on the left-hand side implies the conditional test of ME, given that the OQS[$f$] model holds.
On the other hand, from Theorem \ref{thm3}, an unconditional test of ME is asymptotically equivalent to the conditional test of ME, namely,
\[
	G^{2}({\rm ME})=G^{2}({\rm ME}|{\rm OQS}[f])+o_{p}(1).
\]

\section{Empirical studies}
\label{empirical}
This section presents empirical evaluations of the OQS$[f]$ model through two real-world data analyses. 
Sections \ref{example1} and \ref{example2} demonstrate the model's application to medical and public opinion datasets respectively.
As the OQS$[f]$ model belongs to the multinomial-poisson homogeneous (MPH) model family proposed by \cite{lang2005Homogeneous}, we utilize the \texttt{mph.fit} function in \texttt{R} to implement and analyze the OQS$[f]$ model, facilitating maximum likelihood estimation and comparison with other models.

\subsection{Example 1}
\label{example1}
We consider the data in Table \ref{tab:1} and set equal interval scores $u_{i}=i$ ($i=1,2,3$) for this analysis. 
Table \ref{table:result1} presents the values of the goodness-of-fit test statistics for the applied models. 
We include both the likelihood ratio statistic $G^2$ and the Pearson $X^2$ statistic for each model. 
The Pearson's chi-squared statistic for testing the goodness-of-fit is defined as:
\begin{equation*}
X^{2} = \sum_{i_{1}=1}^{r} \dots \sum_{i_{T}=1}^{r}  \frac{(n_{i_{1}\dots i_{T}} - \hat{m}_{i_{1}\dots i_{T}})^2}{\hat{m}_{i_{1}\dots i_{T}}}.
\end{equation*}
It's important to note that because the table is sparse, we put more faith in the Pearson $X^2$ statistic, using the likelihood ratio statistic $G^2$ mainly to compare models, as suggested by \citet[p.246]{agresti2010Analysis}. 
This approach is particularly relevant for our analysis, given the small sample size ($n = 86$) in this $3 \times 3 \times 3$ contingency table.

\begin{table}[htbp]
    \centering
    \caption{The values of the goodness-of-fit test statistics for models applied to the data in Table \ref{tab:1} \label{table:result1}}
    \vspace{2mm}
    \begin{tabular}{cccc}
        \toprule
        \multirow{2}{*}{Model}  & Deviance & Pearson & Degrees of \\
                            & $G^{2}$ & $X^2$ & freedom \\
        \midrule
        S & 69.00* & 66.21* & 17 \\
        OQS  & \hspace{-2.6mm} 10.35 & \hspace{-2.6mm} 10.01 & 15 \\
        POQS & \hspace{-2.6mm} 24.19 & \hspace{-2.6mm} 22.77 & 15 \\
        MH & 46.28* & 37.88* & 4 \\
        ME & 44.81* & 35.28* & 2 \\
        \bottomrule \vspace{-1mm} \\
        \multicolumn{3}{l}{*Significant at the 5\% level.}
    \end{tabular}
\end{table}

This conclusion is supported by both the Pearson $X^2$ and likelihood ratio $G^2$ statistics, which lead to the same decisions for these cases.
The OQS and POQS models fit well despite having only two more parameters than the S model.
This indicates that these models capture some asymmetry in the data that the S model fails to account for.
Notably, both $X^2$ and $G^2$ statistics agree on the acceptance of these models, reinforcing our confidence in these findings despite the potential instability of $G^2$ in sparse data.
From Theorem \ref{thm2}, we can infer that the poor fit of the S model is primarily due to the poor fit of the ME model rather than the OQS$[f]$ model.

Under the POQS model (setting $\beta_{1}=0$ without loss of generality), we estimate $\hat{\beta}_{2} = 2.30$ and $\hat{\beta}_{3} = 3.35$.
It should be noted that the likelihood ratio test of $H_{0}:\beta_{1}=\beta_{2}=\beta_{3}$ yields $G^{2}(\text{S}|\text{POQS})= 44.81$ with 2 df, rejecting the null hypothesis at the 5\% significance level.
The POQS model reveals the following structures:
\begin{align}
	\hat{\pi}^{c}_{ijk}-\hat{\pi}^{c}_{jik} &= (j - i) \times \frac{2.30}{2\times\#A((i,j,k))} \quad (i<j), \label{B}\\
	\hat{\pi}^{c}_{ijk}-\hat{\pi}^{c}_{kji} &= (k - i) \times \frac{3.35}{2\times\#A((i,j,k))} \quad (i<k), \label{C}\\
	\hat{\pi}^{c}_{ijk}-\hat{\pi}^{c}_{ikj} &= (k - j) \times  \frac{1.05}{2\times\#A((i,j,k))} \quad (j<k). \label{A}
\end{align}
The equations \eqref{B} and \eqref{C} imply that treatments (B) and (C) differ from placebo (A).
The equation \eqref{A} implies that there is only weak evidence that the high dose is better than the low dose.
The analysis of Table \ref{tab:1} using the OQS model is given in \citet{agresti2010Analysis}.
In that case, the asymmetric structure of the odds is focused on.
On the other hand, we focus on the structure of the difference between two conditional probabilities.
The substantive results are the same for the data in Table \ref{tab:1}, despite the different approaches.

We consider the asymptotic property stated in Theorem \ref{thm3}, which claims that $G^2 (\text{S})$ is asymptotically equivalent to the sum of $G^2 (\text{OQS}[f])$ and $G^2(\text{ME})$ under the S model.
The observed values show a substantial discrepancy: while $G^2 (\text{S}) = 69.00$, the sum of $G^2 (\text{OQS}) = 10.35$ and $G^2(\text{ME}) = 44.81$ is only $55.16$, which is less than $G^2(\text{S})$.
This difference illustrates that the asymptotic property is an approximation that improves with larger sample sizes, and deviations are expected in small samples or sparse tables.
Similar trends were observed for other OQS$[f]$ models we examined, specifically those based on Hellinger distance, reverse Kullback-Leibler divergence, and Jensen-Shannon divergence. 
This suggests that the departure from the asymptotic property in small samples is not unique to the OQS model but may be a general characteristic of models within this class.

In contrast, when using the POQS model, we observe a different pattern. 
While $G^2 (\text{S}) = 69.00$ remains the same, we find that $G^2 (\text{POQS}) = 24.19$ and $G^2 (\text{ME}) = 44.81$. 
In this case, the sum $G^2 (\text{POQS}) + G^2 (\text{ME}) = 69.00$ is equal to $G^2 (\text{S})$.
However, it is important to note that Theorem \ref{thm3} establishes an asymptotic equivalence based on large sample theory. 
In Section \ref{asymbehav}, we conducted simulations to examine the asymptotic behavior of these models under varying sample sizes and degrees of asymmetry.

\subsection{Example 2}
\label{example2}
To further explore the performance of the proposed OQS$[f]$ model, we analyze the data in Table \ref{tab:2}, which presents the cross-classification of subjects' opinions on government spending across three domains: Education ($X_1$), Environment ($X_2$), and Assistance to the poor ($X_3$). 
This data, extracted from the 2023 General Social Survey, reflects opinions gathered in 2018 and provides a larger sample ($n = 1065$) in a $3 \times 3 \times 3$ contingency table.
For each variable, responses are categorized as (1) too little, (2) about right, and (3) too much. 
This structure, which includes a neutral response category, might initially suggest equally spaced intervals on a latent scale, potentially justifying the use of equally spaced scoring. 
Based on this assumption, we apply equal interval scores $u_i = i$ for $i=1,2,3$.

Table \ref{table:result2} presents the values of the goodness-of-fit test statistics for the applied models using equal interval scores.

\begin{table}[htbp]
    \centering
    \caption{Opinions about government spending in 2018 from the 2023 General Social Survey.\label{tab:2}}
    \vspace{2mm}    
    \begin{tabular}{ccccccccccc}
	\toprule
		&		&  \multicolumn{3}{c}{$X_{2} = 1$} & \multicolumn{3}{c}{$X_{2} = 2$} & \multicolumn{3}{c}{$X_{2} = 3$}	\\
	\cmidrule(lr){3-11}
	$X_{1}$	&	$X_{3}$	&	1	&	2	&	3	&	1	&	2	&	3	&	1	&	2	&	3	\\
	\midrule
	1      	&           & 558   &  79   &  19   &  95   &  38   &  12   &  21   &   4   &  11  \\
	2      	&           &  38	&  20	&	3	&  37	&  33	&   8	&	4	&  13	&	4	\\
	3      	&           &  14	&	2	&	6	&	7	&	9	&	5	&	9	&	5	&  11    \\
	\bottomrule
    \end{tabular}
\end{table}

\begin{table}[htbp]
    \centering
    \caption{The values of the goodness-of-fit test statistics for models applied to the data in Table \ref{tab:2} \label{table:result2}}	
    \vspace{2mm}
    \begin{tabular}{ccc}
        \toprule
        \multirow{2}{*}{Model}  & Deviance  & Degrees of \\
                                & $G^{2}$   & freedom \\
        \midrule
        S       & 49.47* & 17 \\
        OQS     & \hspace{-2.6mm} 24.32 & 15 \\
        POQS    & \hspace{-2.6mm} 23.40 & 15 \\
        MH      & 38.21* & 4 \\
        ME      & 26.07* & 2 \\
        \bottomrule
        \multicolumn{3}{l}{*Significant at the 5\% level.}
    \end{tabular}
\end{table}

We estimate $\hat{\beta}_{2} = 1.78$ and $\hat{\beta}_{3} = 0.95$ under the POQS model (setting $\beta_{1}=0$ without loss of generality).
The likelihood ratio test of $H_{0}:\beta_{1}=\beta_{2}=\beta_{3}$ yields $G^{2}(\text{S}|\text{POQS})= 26.07$ with 2 df, rejecting the null hypothesis at the 5\% significance level.
Then, the POQS model reveals the following structures:
\begin{align}
	\hat{\pi}^{c}_{ijk}-\hat{\pi}^{c}_{jik} &= (j - i) \times \frac{1.78}{2\times\#A((i,j,k))} \quad (i<j), \label{B2}\\
	\hat{\pi}^{c}_{ijk}-\hat{\pi}^{c}_{kji} &= (k - i) \times \frac{0.95}{2\times\#A((i,j,k))} \quad (i<k), \label{C2}\\
	\hat{\pi}^{c}_{ijk}-\hat{\pi}^{c}_{ikj} &= (k - j) \times  \frac{-0.83}{2\times\#A((i,j,k))} \quad (j<k). \label{A2}
\end{align}
The equations \eqref{B2} and \eqref{C2} imply that subjects tend to perceive government spending on Environment and Assistance to the poor as being greater than spending on Education.
The equation \eqref{A2} implies that subjects tend to view government spending on the Environment as being higher than on Assistance to the poor.

The results from this larger sample ($n = 1098$) demonstrate a closer adherence to the asymptotic property stated in Theorem \ref{thm3}, compared to the significant violation observed in Section \ref{example1} with a smaller sample size ($n = 86$).
Unlike the results from Section \ref{example1}, where the asymptotic property stated in Theorem \ref{thm3} was significantly violated, the increased sample size in this example appears to yield results closer to the expected asymptotic behavior. 
The difference between $G^2(\text{S})$ and the sum of $G^2(\text{OQS})$ and $G^2(\text{ME})$ is considerably smaller compared to that observed in Section \ref{example1}, suggesting that the asymptotic property described in Theorem \ref{thm3} is more closely approximated in this larger sample.

While the S model still does not provide a statistically significant fit ($G^2 = 49.47$, df = 17, $p < 0.05$), its fit has improved compared to that of Section \ref{example1} ($G^2 = 69.00$, df = 17, $p < 0.05$).
These observations suggest that the improved adherence to the asymptotic property may be influenced not only by the increased sample size but also by the specific characteristics of this dataset. 
The complex interplay between sample size, data structure, and model performance underscores the need for further research to fully understand the behavior of these models under various conditions. 
In Section \ref{asymbehav}, we provides a systematic investigation of how introducing asymmetric structures to a symmetric base affects model performance across varying sample sizes.

\section{Simulation studies}
\label{simulation}
This section presents comprehensive simulation studies to evaluate the proposed model's limitations and asymptotic properties.
Section \ref{paraboot} employs parametric bootstrap analysis to assess the statistics' behavior with small sample sizes, using the data from Table \ref{tab:1}.
Section \ref{normal} examines the empirical power across several parameter configurations for multivariate normal data.
Section \ref{asymbehav} investigates the asymptotic properties stated in Theorem \ref{thm3} by examining test statistic relationships under several asymmetric structures.

\subsection{Parametric bootstrap for small samples}
\label{paraboot}
Since the sample size in Table \ref{tab:1} is small, with only 86 observations distributed across 27 cells, we consider methods that do not rely solely on the chi-squared approximation for model evaluation. 
Specifically, we employ parametric bootstrap methods to assess the performance of $G^{2}$ in this sparse data.
The parametric bootstrap procedure for a model proceeds as follows. 
\begin{enumerate}[leftmargin=*, label=Step \arabic*.] \itemsep=2mm
\item Compute the MLEs of $\bm{\pi}$, $\hat{\bm{\pi}}$, under model M for the data in Table \ref{tab:1}.
\item Compute the value of $G^{2}(\text{M})$.
\item Generate $10^{5}$ multinomial samples with parameters (86, $\hat{\bm{\pi}}$), that is, contingency tables.
\item For each generated contingency table, compute the MLEs $\hat{\bm{\pi}}_{s}$ and $G^{2}_{s}(\text{M})$ $(s=1,\dots,10^{5})$. 
\item Compare $G^{2}(\text{M})$ with the simulated values to obtain the $p$-value and the 0.05 critical value.
\end{enumerate}

The results of the parametric bootstrap analysis for the various models are as follows. For the S model, the obtained $p$-value is 0.000, and the 0.05 critical value is 29.52, while the chi-squared critical value with 17 df is 27.59. 
This indicates that there is a relatively small difference between the two critical values. 
For the OQS and POQS models, the $p$-values are 0.874 and 0.079, respectively, with corresponding 0.05 critical values of 24.72 and 25.80. 
The chi-squared critical value with 15 df is 25.00, suggesting a small but noticeable difference between the critical values for these models.

In contrast, for the ME and MH models, the $p$-values are 0.002 and 0.001, respectively, and their 0.05 critical values are 33.74 and 33.48. 
The chi-squared critical values for these models are 5.99 with 2 df and 9.49 with 4 df, respectively, indicating a larger discrepancy between the parametric bootstrap and chi-squared test results.
On the other hand, for $G^{2}(\text{S}|\text{OQS})$ and $G^{2}(\text{S}|\text{POQS})$, the $p$-values are 0.000 and 0.000, respectively, and their 0.05 critical values are 8.13 and 6.60. 
The chi-squared critical value is 5.99 with 2 df.
These results may indicate that the conditional test performs better than the unconditional test in the parametric bootstrap method.

In conclusion, the parametric bootstrap results generally align with the chi-squared test results for certain models. 
However, notable differences in critical values were observed in some cases, particularly for the ME and MH models. 
This aligns with previous findings in the literature regarding the challenges of using goodness-of-fit tests in sparse data scenarios.
\cite{vondavier1997Bootstrapping} demonstrated through extensive Monte Carlo simulations that $G^2$ can be unstable for sparse data, potentially leading to the incorrect acceptance of underfitting models. 
Their study showed that while bootstrap methods are generally effective for sparse categorical data, Pearson's $X^2$ statistic outperforms $G^2$, especially in sparse conditions. 
Similarly, \cite{reiser2019Goodnessoffit} highlighted the difficulties of using traditional goodness-of-fit test statistics like $G^2$ in high-dimensional and sparse contingency tables. 
They found that bootstrapped methods can mitigate some issues, but $G^2$ remains sensitive to sparseness, often leading to misleading conclusions.

Given these limitations of both chi-squared approximations and parametric bootstrap methods for sparse data, future work should explore more precise exact testing methods. 
Additionally, following Agresti's recommendation, it may be prudent to rely more heavily on the Pearson $X^2$ statistic for goodness-of-fit assessments in sparse tables like the one in our study.

\subsection{Model performance under multivariate normality}
\label{normal}
For two-way contingency tables, \cite{agresti1983Simple} pointed out that the LDPS model, which is equivalent to the proposed model with equally spaced integer scores $\{u_{i}=i\}$, is adequate for the table underlying bivariate normal distribution with different means and same variances. 
Extending this concept, \cite{tahata2020Separation} conducted extensive simulation studies using $6 \times 6$ tables to investigate the performance of models based on Pearsonian distance, which are extensions of the LDPS model using $f$-divergence.

For multi-way contingency tables, \cite{yamamoto2007Decomposition} proposed a generalization of the LDPS model to three-way tables. 
They examined the model's performance using $4 \times 4 \times 4$ contingency tables generated from underlying trivariate normal distributions with various means, variances and correlation coefficients. 
However, their analysis was limited in scope, as they evaluated the goodness-of-fit for only one contingency table (with 10,000 observations) per parameter configuration, rather than assessing the empirical power through repeated simulations.

Our primary goal is to assess the empirical power of the proposed model in detecting asymmetric structures in these multidimensional relationships and to what extent it maintains its performance when applied to higher-dimensional data. 
By comparing the model's performance across different parameter configurations, we aim to gain valuable insights into its effectiveness in identifying various types of asymmetric structures in higher-dimensional settings. 
This extension to higher dimensions and more thorough simulation is particularly important because the complex dependencies present in multivariate distributions may not allow for a straightforward generalization of properties observed in the bivariate case. 
Through this study, we seek to understand how well the proposed model can discern different patterns of asymmetric structures and how its detection capability changes as we transition from bivariate to multivariate frameworks.

In this simulation study, we assume random sampling from an underlying trivariate normal distribution characterized by mean vector $(\mu_{1}, \mu_{2}, \mu_{3})$, variance vector $(\sigma_{1}^{2}, \sigma_{2}^{2}, \sigma_{3}^{2})$, and correlation coefficient vector $(\rho_{12}, \rho_{13}, \rho_{23})$. 
We consider $12$ different scenarios, representing all combinations of the following parameters:
\begin{align*}
    (\mu_{1}, \mu_{2}, \mu_{3})                         &= (0.0, 0.0, 0.0), (0.0, 0.0, 0.1), (-0.1, 0.0, 0.1), \\
    (\sigma_{1}^{2}, \sigma_{2}^{2}, \sigma_{3}^{2})    &= (1.0, 1.0, 1.0), (1.0, 1.0, 1.2), \\
    \text{and} \hspace{2mm} (\rho_{12}, \rho_{13}, \rho_{23})                   &= (0.2, 0.2, 0.2), (0.2, 0.2, 0.3).
\end{align*}
Following the approach of \cite{yamamoto2007Decomposition}, we generate $4 \times 4 \times 4$ contingency tables using identical boundaries for all three variables at $\mu_{2}$, $\mu_{1} \pm 0.6 \sigma_{1}$. 
The simulation procedure is as follows.
\begin{enumerate}[leftmargin=*, label=Step \arabic*.] \itemsep=2mm
\item Generate 10000 random numbers from a trivariate normal distribution with parameters $(\mu_{1}, \mu_{2}, \mu_{3})$, $(\sigma_{1}^{2}, \sigma_{2}^{2}, \sigma_{3}^{2})$ and $(\rho_{12}, \rho_{13}, \rho_{23})$.
\item A $4 \times 4 \times 4$ contingency table is formed using cut points for all three variables at $\mu_2$, $\mu_1 \pm 0.6\sigma_1$.
\item Models are fitted for the table.
\item Steps 1, 2 and 3 are repeated 10,000 times.
\end{enumerate}

Table \ref{tab:3} gives the values of empirical power of each model for 12 different cases.

\begin{table}[htbp]
  \caption{Empirical power of each model for trivariate normal distributions across different parameter configurations}
  \vspace{2mm}
  \label{tab:3}
  \centering
  \begin{tabular}{cccccc}
   \toprule
    &  &  & \multicolumn{3}{c}{Models} \\
   \cmidrule(lr){4-6}
    ($\mu_{1}, \ \mu_{2}, \ \mu_{3}$) & ($\sigma_{1}^{2}, \ \sigma_{2}^{2}, \ \sigma_{3}^{2}$) & ($\rho_{12}, \ \rho_{13}, \ \rho_{23}$) & S & OQS & POQS \\
   \midrule
   ($0.0, 0.0, 0.0$) & ($1.0, 1.0, 1.0$) & ($0.2, 0.2, 0.2$) & $0.0474$ & $0.0514$ & $0.0508$ \\
   ($0.0, 0.0, 0.0$) & ($1.0, 1.0, 1.0$) & ($0.2, 0.2, 0.3$) & $0.9991$ & $0.9992$ & $0.9991$ \\
   ($0.0, 0.0, 0.0$) & ($1.0, 1.0, 1.2$) & ($0.2, 0.2, 0.2$) & $0.8947$ & $0.9007$ & $0.8995$ \\
   ($0.0, 0.0, 0.0$) & ($1.0, 1.0, 1.2$) & ($0.2, 0.2, 0.3$) & $1.0000$ & $1.0000$ & $1.0000$ \\ \\

   ($0.0, 0.0, 0.1$) & ($1.0, 1.0, 1.0$) & ($0.2, 0.2, 0.2$) & $0.9993$ & $0.0585$ & $0.0603$ \\
   ($0.0, 0.0, 0.1$) & ($1.0, 1.0, 1.0$) & ($0.2, 0.2, 0.3$) & $1.0000$ & $0.9991$ & $0.9980$ \\
   ($0.0, 0.0, 0.1$) & ($1.0, 1.0, 1.2$) & ($0.2, 0.2, 0.2$) & $1.0000$ & $0.9142$ & $0.9276$ \\
   ($0.0, 0.0, 0.1$) & ($1.0, 1.0, 1.2$) & ($0.2, 0.2, 0.3$) & $1.0000$ & $1.0000$ & $1.0000$ \\ \\

   ($-0.1, 0.0, 0.1$) & ($1.0, 1.0, 1.0$) & ($0.2, 0.2, 0.2$) & $1.0000$ & $0.0770$ & $0.0920$ \\
   ($-0.1, 0.0, 0.1$) & ($1.0, 1.0, 1.0$) & ($0.2, 0.2, 0.3$) & $1.0000$ & $0.9985$ & $0.9994$ \\
   ($-0.1, 0.0, 0.1$) & ($1.0, 1.0, 1.2$) & ($0.2, 0.2, 0.2$) & $1.0000$ & $0.8710$ & $0.8981$ \\
   ($-0.1, 0.0, 0.1$) & ($1.0, 1.0, 1.2$) & ($0.2, 0.2, 0.3$) & $1.0000$ & $1.0000$ & $1.0000$ \\
   \bottomrule
  \end{tabular}
\end{table}

The simulation results in Table \ref{tab:3} demonstrate that both OQS and POQS models exhibit excellent performance when the underlying trivariate normal distribution has equal variances and correlation coefficients, regardless of mean differences. 
This finding extends the properties in two-way contingency tables argued by \cite{agresti1983Simple} to the multi-way framework.
It confirms that our proposed models primarily capture probability structures allowing for mean differences while assuming homogeneity in other parameters.
In our analysis of various scenarios, we observed that both OQS and POQS models exhibit similar capabilities in detecting asymmetric structures. 
Further investigations, not shown in the table, revealed that models based on Hellinger distance, reverse Kullback-Leibler divergence, and Jensen-Shannon divergence also demonstrate comparable performance.
This similarity in performance across models based on different divergence measures indicates that the choice of measure does not substantially affect the models' ability to identify asymmetric structures.
As a result, analysts may have flexibility in choosing a divergence measure when expressing relationships between cell probabilities of interest in multi-way contingency tables.

Our extensive simulations, showing high empirical powers, provide statistically rigorous verification of the results in \cite{yamamoto2007Decomposition}, demonstrating the models' capability in detecting asymmetries in variances and correlation coefficients.
These findings extend the properties from two-way to multi-way tables and offer valuable insights for researchers analyzing multi-way contingency tables derived from multivariate normal data.

\subsection{Asymptotic equivalence for asymmetric structure}
\label{asymbehav}
To validate the asymptotic equivalence stated in Theorem \ref{thm3} and explore its behavior under varying asymmetric structures, we conducted a simulation study.
Our aim is to empirically demonstrate the convergence of the relationships between likelihood ratio statistics for the S, OQS$[f]$, and ME models across different sample sizes and degrees of asymmetry. 
We designed our simulation based on $3 \times 3 \times 3$ contingency tables, introducing varying degrees of asymmetry to a uniform distribution.
For this simulation, we used equally spaced integer scores ${u_{i}=i}$ for both the OQS$[f]$ and ME models, following common practice in ordinal data analysis.
Our simulation procedure gradually shifts the probability structure from complete symmetry to asymmetry, allowing us to observe the behavior across this transition, as follows.
\begin{enumerate}[leftmargin=*, label=Step \arabic*.] \itemsep=2mm
    \item Generate a $3 \times 3 \times 3$ contingency table with an underlying uniform distribution, where the cell probabilities are all $1/27$.
    \item For a fixed degree of asymmetry $d$, and each off-diagonal cell $\boldsymbol{i}$ with ascending order : \vspace{1mm}
    \begin{enumerate} \itemsep=2mm
        \item Increase the probability of cell $\boldsymbol{i}$ by $d$.
        \item Decrease the probabilities of all other cells in $D(\boldsymbol{i})$ by $d/(|D(\boldsymbol{i})|-1)$.
    \end{enumerate}
    \item Set this as the true probability distribution and generate $10000$ tables for each sample size $n = 250, 500, 1000,2000,\dots,20000$.
    \item For each sample size $n$, compute the mean squared error (MSE) :
    \begin{equation*}
        \frac{1}{10000} \sum_{s=1}^{10000} \left(\frac{G_s^2(\text{S})}{G_s^2(\text{OQS}[f]) + G_s^2(\text{ME})} - 1\right)^2.
    \end{equation*}
    where $G_{s}^{2}$ is the likelihood ratio statistic calculated for the $s$-th table.
    \item Repeat Steps 1-4 for each $d = 0, 0.005, 0.01, 0.02, 0.03$.
\end{enumerate}

\begin{figure}[htbp]
    \centering
    \includegraphics[width=13cm]{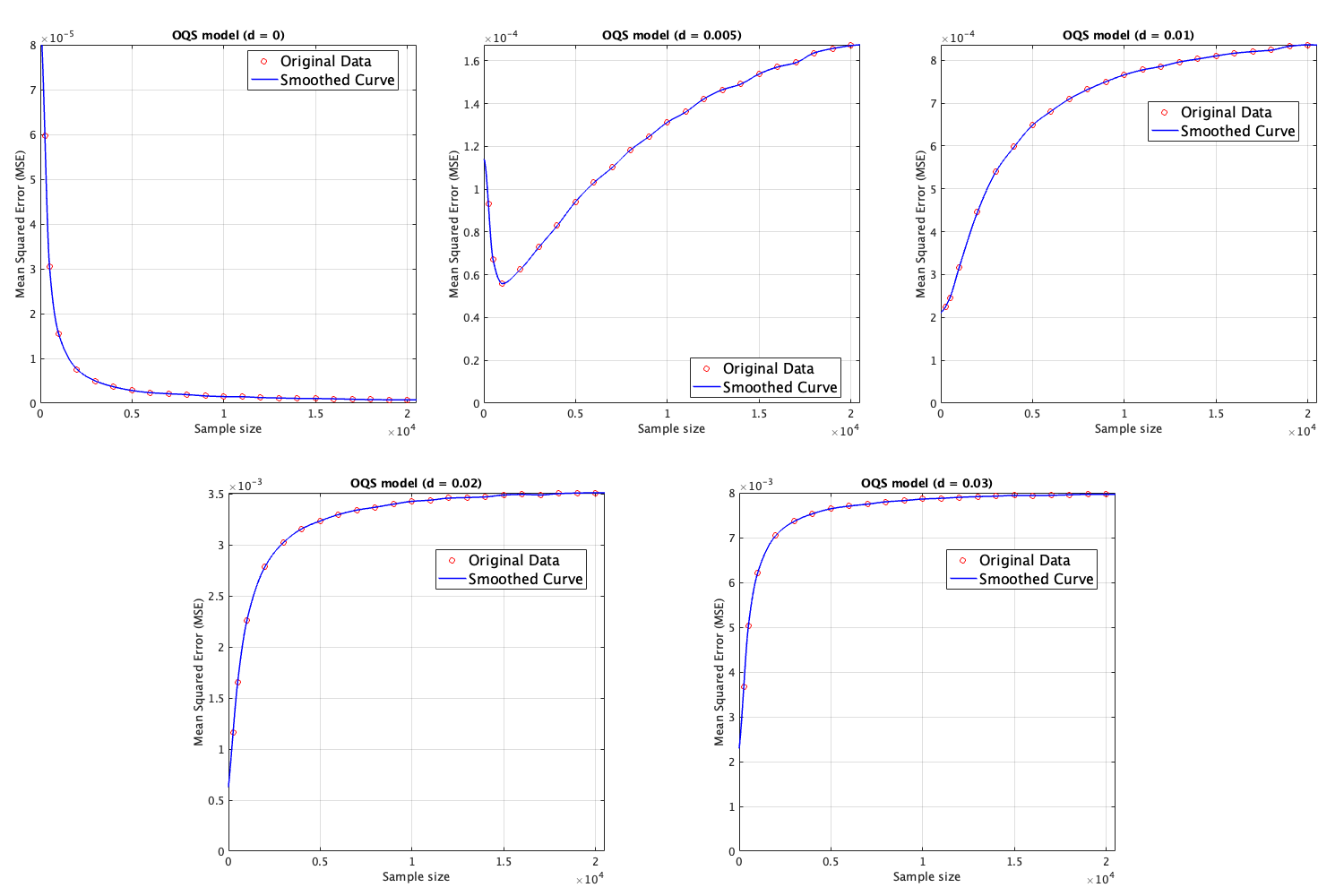}
    \caption[]{Asymptotic behavior of MSE for OQS model with varying degrees of asymmetry \label{fig:convergence}}
\end{figure}

We focused our investigation on the OQS and POQS models, examining their behavior across different degrees of asymmetry $d$ and sample size $n$.
Figure \ref{fig:convergence} illustrates the behavior of the MSE for the OQS model as the $n$ increases, for each $d$.
It is important to note that to ensure computational stability, particularly for smaller sample sizes, we added $1/2$ to each observed frequency in the contingency tables. 
This standard adjustment technique helps prevent issues with zero cells without significantly altering the underlying distribution structure. 

For the OQS model, we observe distinct patterns across different asymmetry levels: 
\begin{itemize} \itemsep=2mm
    \item Under complete symmetry ($d=0$), the MSE decreases rapidly as $n$ increases up to 2000, converging asymptotically to zero. 
    This behavior aligns with the asymptotic equivalence derived from Wald statistics in Theorem \ref{thm3}.
    
    \item With slight asymmetry ($d=0.005$), the MSE initially decreases for smaller sample sizes ($n = 250, 500, 1000$) but increases thereafter. 
    This trend likely reflects the growing statistical power to detect departures from complete symmetry as $n$ increases.
    
    \item For moderate to high asymmetry ($d=0.01, 0.02, 0.03$), the MSE is substantially larger compared to the above cases $d=0, 0.005$, and consistently increases with sample size.
    Notably, for $d=0.01$, the MSE at $n = 250$ is slightly elevated, possibly due to limited power at small sample sizes. 
\end{itemize}
These results provide empirical support for Theorem \ref{thm3}, demonstrating that the asymptotic equivalence holds under symmetric structure.
This finding is consistent with the known asymptotic equivalence of Wald and likelihood ratio statistics under the null hypothesis \cite[Sec. 6e. 3]{rao1973Linear}.

In contrast, the POQS model exhibits remarkably consistent behavior across all degrees of asymmetry. 
The MSE values for this model remain exceptionally low, approximately $10^{-18}$, regardless of the asymmetry level or sample size. 
While Theorem \ref{thm3} theoretically applies only under symmetry, the POQS model appears to maintain this property across all examined scenarios. 

\begin{figure}[htbp]
    \centering
    \includegraphics[width=13cm]{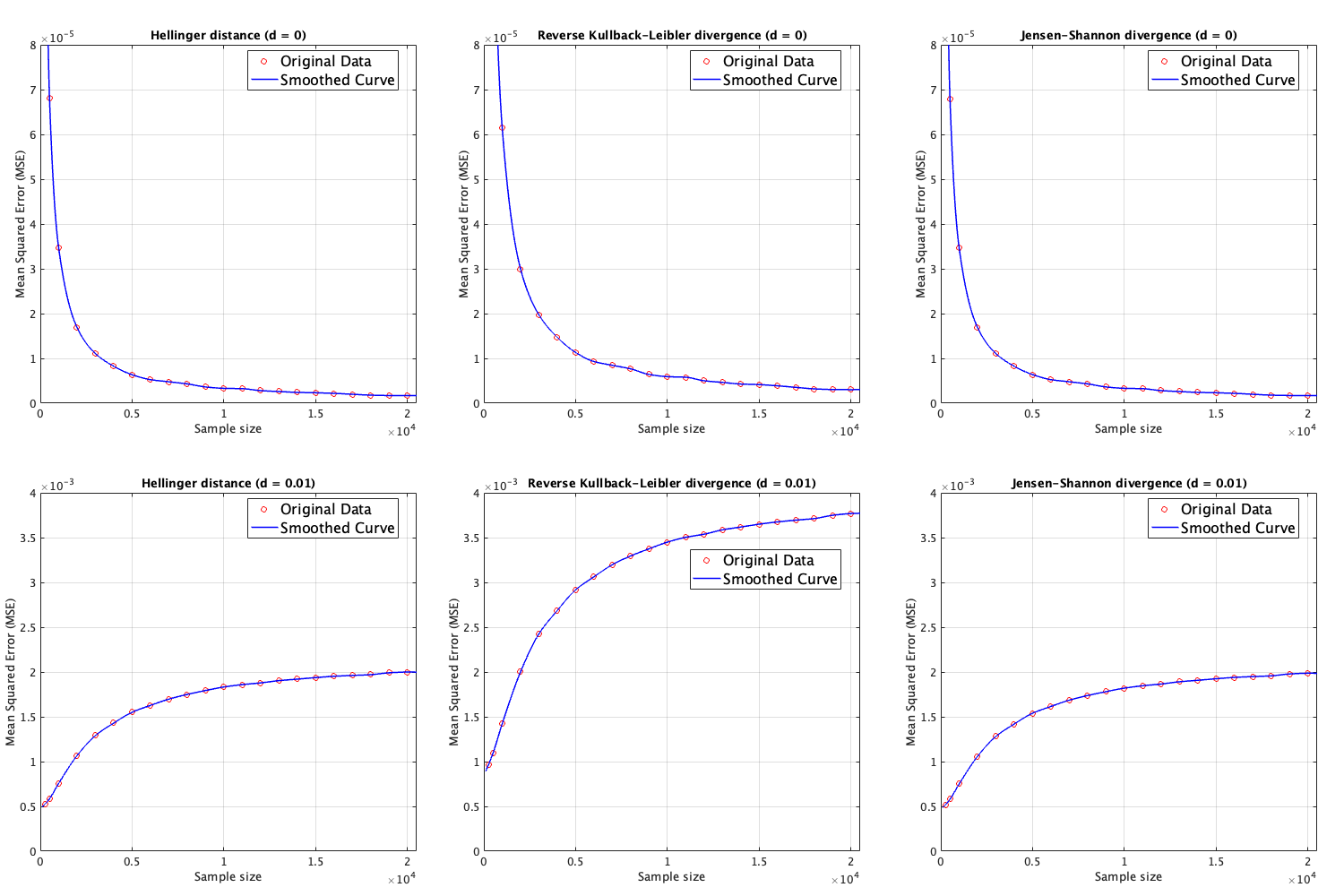}
    \caption[]{Asymptotic behavior of MSE for OQS$[f]$ models based on Hellinger distance, reverse Kullback-Leibler divergence, and Jensen-Shannon divergence, with varying degrees of asymmetry \label{fig:otherfunc}}
\end{figure}

To further investigate the generality of these findings, we extended our analysis to the proposed models based on Hellinger distance, reverse Kullback-Leibler divergence, and Jensen-Shannon divergence. 
We focused our simulation on two key scenarios: complete symmetry ($d=0$) and moderate asymmetry ($d=0.01$). 
As shown in Figure \ref{fig:otherfunc}, these models exhibit behavior similar to the OQS model in both scenarios. 
Under complete symmetry, all models show a rapid decrease in MSE as the $n$ increases, while under moderate asymmetry, they consistently demonstrate an increasing trend in MSE with larger $n$. 
The similar behavior to the OQS model observed across various divergence measures and asymmetry conditions emphasizes the distinctive characteristics of the POQS model. 
These results not only validate the asymptotic properties stated in Theorem \ref{thm3} but also reveal unexpected behaviors, particularly for the POQS model.

While most of the examined OQS$[f]$ models behave as expected under asymptotic theory with increasing MSE as asymmetry grows, the POQS model demonstrates remarkably stable performance across all asymmetry conditions. 

\section{Concluding remarks with discussions}
\label{fin}
In this paper, we have accomplished three objectives.
Firstly, we have proposed the OQS[$f$] model.
The model is based on the $f$-divergence and allows us to consider various asymmetry models by selecting the function $f$.
When the S model is not applicable to a given data, these models may fit the data well without additional parameters.
Secondly, we have outlined the properties of the OQS[$f$] model and provided an information-theoretic interpretation of the model from Theorem \ref{thm1}.
Lastly, we revisited Agresti's findings and presented a new necessary and sufficient condition for the S model.
The OQS[$f$] and ME models are the separable hypotheses from Theorems \ref{thm2} and \ref{thm3}.

Consider again the data in Table \ref{tab:1}.
The MH model fits the data poorly from Table \ref{tab:2}.
In this case, a model with weaker restrictions than the MH model will be applied.
As such a model, the marginal logistic model (ML) model is considered in, for example, \citet[p.241]{agresti2010Analysis}.
The ML model is defined as follows:
\[
	\log \frac{F_{i}^{(t)}}{{1-F_{i}^{(t)}}}=\log \frac{F_{i}^{(1)}}{{1-F_{i}^{(1)}}}-\delta_{t-1} \quad (i=1,\dots,r-1; t=2,\dots,T),
\]
where $F_{i}^{(t)} = \Pr(X_{t} \leq i)$.
The value of $G^{2}(\rm{ML})$ is $0.52$ with 2 df.
The fit is quite good.
The estimated marginal cumulative distributions under the ML model are
\[
	\hat{F}_{i}^{(2)}=\frac{\exp(\hat{\theta}_{i}-2.04)}{1+\exp(\hat{\theta}_{i}-2.04)} \quad \mbox{and} \quad \hat{F}_{i}^{(3)}=\frac{\exp(\hat{\theta}_{i}-2.43)}{1+\exp(\hat{\theta}_{i}-2.43)},
\]
where $\hat{\theta}_{i} = \log (\hat{F}_{i}^{(1)}/(1-\hat{F}_{i}^{(1)}))$.
We can achieve comparable outcomes using the POQS (or OQS) model.
The POQS (or OQS) model can provide a more detailed data analysis than the ML model.
This is because the ML model only shows the structure of the marginal distribution, whereas the POQS (or OQS) model reveals the structure of the joint probabilities.
Furthermore, \citet{tahata2007Decompositions} showed that the MH model holds if and only if the ML and ME models hold for multi-way contingency tables.
However, the separability of ML and ME has yet to be proven.

The proposed OQS$[f]$ model demonstrates significant potential in expressing a wide range of probability structures previously challenging to capture. 
By utilizing $f$-divergence, a broad family of divergence measures including Cressie–Read power divergence \citep{cressie1984Multinomial} as a prominent example, we have expanded our ability to describe diverse probability structures in multi-way contingency tables.
This advancement allows for capturing complex ordinal data, previously undetectable.
However, it is important to acknowledge potential limitations, particularly as dimension and category numbers increase in larger multi-way tables. 
The current parameterization, relying solely on $\{\beta_{i}\}$, may face constraints in expressing complex relationships beyond linear associations.
To address these challenges and enhance the model's applicability, future work could focus on introducing more flexible, yet interpretable parameters. 
This might involve exploring non-linear relationships or incorporating additional parameters to capture more complex interactions, aiming to maintain interpretability while expanding capability for complex ordinal data structures.

An important area for future research is the exploration of different scoring approaches for our model. 
While we used equal interval scores, studies like \cite{bradley2015Rating} and \cite{lantz2013Equidistance} have highlighted the complexities in score determination for ordinal data, including challenges with neutral categories and the validity of equal intervals. 
Additionally, the types of scores for ordinal categories are described in \cite{chen2014Assignment} and \citet[p.9]{agresti2010Analysis}.
Future work should investigate how alternative scoring methods (e.g., non-integer, nonlinear, or data-driven) affect the model's performance and properties. 
We recommend researchers carefully examine their data characteristics and explore various scoring options. 
Our proposed framework is flexible enough to accommodate different scoring systems, potentially enhancing its applicability across diverse ordinal data. 
This adaptability allows researchers to tailor their analysis to the specific requirements of their data, though detailed investigations into optimal scoring strategies remain a task for future research.

The \texttt{mph.fit} function essential for fitting our proposed models can be obtained directly from its developer.
For researchers interested in applying our OQS$[f]$ model to their own data, we are able to provide the necessary code for implementing the specific parameterization and constraints required for our model. 
This includes the code for setting up the model structure and defining constraint conditions within the \texttt{mph.fit} framework. 
We encourage interested researchers to contact the corresponding author for access to this supplementary code and guidance on its use with the \texttt{mph.fit} function.

\section*{Acknowledgements}
This work was supported by JSPS KAKENHI Grant Number JP20K03756. 
The authors express gratitude to Prof. Joseph B. Lang for generously sharing his R program, which was instrumental in conducting the analysis presented in the example.

\bibliographystyle{apalike}
\bibliography{arXiv.bib}

\end{document}